\documentclass[twocolumn,showpacs,preprintnumbers,amsmath,amssymb]{revtex4}

\usepackage{graphicx}
\usepackage{dcolumn}
\usepackage{bm}

\begin{document}

\preprint{APS/123-QED}

\title{Concentration-Temperature Superposition of Helix Folding
Rates in Gelatin}

\author{J. L. Gornall,  E. M. Terentjev}
\affiliation{Cavendish Laboratory, University of Cambridge, J J
Thomson Avenue, Cambridge, CB3 0HE, U.K.}

\date{\today}

\begin{abstract}
We study the kinetics of helix-coil transition in water solutions
of gelatin (collagen protein) by optical rotation techniques
combined with thermal characterization. By examining the rates of
secondary helix folding, and covering a very wide range of
solution concentrations, we are able to identify a universal
exponential dependence of folding rate on concentration and quench
temperature. We demonstrate a new concentration-temperature
superposition of data at all temperatures and concentrations, and
build the corresponding master curve. The results support the
concept of a diffuse helix-coil transition.  We find no
concentration dependance of the normalized rate constant,
suggesting first order (single) kinetics of secondary helix
folding dominate in the early stages of renaturation.
\end{abstract}


\maketitle

Kinetics of protein folding is one of the cornerstone problems in
understanding the full mystery of biologically active macromolecules
\cite{fersht03,kubel04}. A number of experimental techniques focus
on a great variety of natural and synthetic polypeptides undergoing
their globular collapse, or denaturation into the coil. Although one
might want to think of such a transition in terms of individual
macromolecules, there is much evidence that cooperative effects
between different chains play a significant role, an effect often
referred to as oligomerization \cite{elco01}. This extreme
variability has, to date, defied any attempt on universal
description, beyond the classical Zimm-Bragg abstraction
\cite{zimm59}.

In this Letter we examine perhaps the most well-studied protein,
collagen, and discover that the complexity of its folding kinetics,
strongly dependent on temperature and concentration in the solvent,
can be dramatically simplified. All of this kinetics can be scaled
onto a single master curve by a new procedure we call the
``concentration-temperature superposition''. There is a remarkable
analogy: the classical time-temperature superposition
\cite{ferry70,wlf55} has allowed master curves to describe the glass
transition in a variety of thermal viscoelastic systems and served a
great purpose in rheology for the last 50 years. A much more recent
discovery of time-concentration superposition \cite{trap00,cicu03}
has allowed the universal description of dynamic glass, or jamming
transition, in lyotropic systems such as colloid suspensions. In
both cases the cooperativity of interparticle interaction is the
key, and we shall argue that it is relevant for the folding kinetics
as well.

We must emphasize that we study the principal helix-coil transition
in collagen chains, i.e. the folding of secondary helices. This is
achieved by our experimental methods focusing on optical rotation of
linearly polarized light at a wavelength away from any absorbtion
band. It has been recently demonstrated that the dominant
contribution to the measured rotation rate arises from the secondary
helices \cite{gornall06}. This is different from the more common
studies of gelation kinetics in gelatin, which is controlled by the
tertiary triple helices and is usually studied by rheological
methods. The field of sol-gel transition and mechanical response of
networks is broad and well-established \cite{winter86,goldbart96}.
Recent important observations~\cite{normand00} of universal master
curves, describing specifically the storage modulus in gelatin, also
make connections with glassy dynamics. We shall see from the results
below that our present work is not related to these ideas: we
examine much shorter times at which the secondary structure is
formed, while the kinetics of subsequent rheological is a much
slower process.

Collagen is the main protein component of white fibrous connective
tissues such as skin, tendons and bones. The fundamental unit of
the native collagen structure is a rod consisting of three
individual molecular strands, each twisted into a secondary
left-handed helix, which analogous and very similar to the
classical $\alpha$-helix in other polypeptides. In collagen, three
segments of secondary helix wrap together to form a tertiary
right-handed superhelix stabilized by further interchain hydrogen
bonding \cite{rama67}.  Collagen is extracted from tissues by
hydrolytic degradation, which denatures it producing the resulting
material commonly known as gelatin.

Gelatin is dissolved in water by heating the solution to
40$^{\circ}$C. Above this temperature collagen chains have random
coil configuration. On cooling transparent gels form containing
extended physical crosslinks. X-ray diffraction and transmission
electron microscopy measurements suggest that the crosslinks are
formed by partial reversion to ordered triple-helical segments,
separated along the chain contour by peptide residues still in the
random coil configuration \cite{pezr90,djab93}.  At very low
concentrations the renaturation, or folding process is completely
intramolecular and proceeds by a back folding of the single chains
\cite{harr70}. With increasing concentration the renaturation
becomes increasingly intermolecular. Gelation occurs at
concentrations above $5$\,mg/ml due to the formation of an
infinite elastic network in the gelatin solution, see
\cite{normand00,guo03} and references therein.

\begin{figure} [b]
\centering \resizebox{0.47\textwidth}{!}{\includegraphics{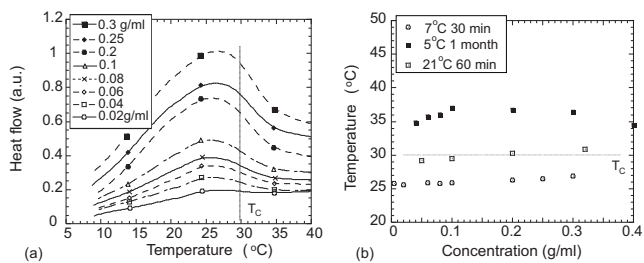}}
\caption{(a) DSC traces demonstrating the helix-coil transition in
gelatin/water solutions of different concentrations, labelled on
plot. \ (b) A thermal phase diagram of this transition, $T_m(c)$,
obtained from the DSC data ($\odot$), with results for different
preparation methods from the literature ($\boxdot$ - \cite{god78},
$\blacksquare$ - \cite{bohi93}). The value $T_c$ is obtained from
the concentration-temperature superposition analysis below.}
\label{fig:phase}
\end{figure}

Gelatin owes many of its uses to this coil-helix transition.  The
kinetics of the transition has been extensively studied for many
years using a variety of techniques such as differential scanning
calorimetry (DSC), scattering and rheometry, as well as optical
rotation. Figure~\ref{fig:phase}(a) shows a sequence of DSC traces
of helix-coil transition that we obtained in gelatin solutions of
different concentration, while Fig.~\ref{fig:phase}(b) shows the
corresponding transition line, $T_m(c)$. A very flat dependence of
the transition temperature on concentration in the thermal phase
diagram, extrapolating to $c \rightarrow 0$, indicates that the
transition is largely intrachain. Note that the gelation phase
diagram, obtained by rheological methods, would show a drop of the
gel point $T_g(c)$ below $5$\,mg/ml. Also note that, since the
melting temperature depends strongly on the gelation temperature and
time, a broad range of $T_m(c)$ values are found in the literature.

Gelatin, as any protein, is an optically active material in both the
random coil and helical states. However, due to coherent long-range
chiral ordering, helical domains rotate the plane of light
polarization much more strongly than the individual chiral
aminoacids in the coil state. Thus, the coherent optical activity
gives a direct indication of the fraction of the monomers in helical
states.

In many biopolymers, including polysaccharides and DNA, the
coil-helix transition is very fast and resembles a true first order
phase transition \cite{nort83}.  In gelatin, however, the helix
nucleation step lasts several minutes and subsequent growth of the
helices proceeds even slower, at a logarithmic rate \cite{djab88}.
This allows one to study its detailed kinetics. Most investigations
have focused on isothermal renaturation rather than temperature
scanning studies, because of the difficulty in accessing the
equilibrium state. This slowness of the process will allow us to
scan and superpose the renaturation kinetics over a range of
temperatures and concentrations.

In one of the earlier studies, Flory and Weaver studied helix growth
rates in very dilute gelatin solutions ($c<4$\,mg/ml) and found
first order kinetics \cite{flor60}, i.e.~that the folding rate is
concentration-independent. They postulated the coil-helix transition
proceeds via an intermediate state formed by intramolecular
rearrangement of a single chain. Assuming this state consists of a
secondary helix segment, consideration of the minimum stable segment
length leads to the Flory-Weaver expression for the rate constant
for the renaturation after quenching the dilute solution:
\begin{equation}
k_1 = B\exp\left({\frac{-A}{kT \Delta{T}}}\right) ,
\label{eq:flory}
\end{equation}
where $A$ and $B$ are constants, $T$ is the quench temperature,
and $\Delta{T}= T_{\rm m} - T$ is the degree of supercooling below
the equilibrium melting temperature $T_{\rm m}$.

The first order kinetic analysis agrees with other experimental
observations of dilute gelatin solutions \cite{harr70}.  In
semidilute solutions concentration-dependant kinetics has been
observed \cite{guo03,term90}. Consequently, new mechanisms have been
proposed, involving either different segments of the same chain or
up to three different peptide chains having to interact to form
stable helices. The topological consequences of these arrangements
on helix formation and gelation are very significant.

Goddard {\it et al.}~\cite{god78}, suggesting an analogy between
renaturation and crystallization, modelled the kinetics in terms of
Avrami exponents \cite{avr41}: $ \chi = 1-\exp \left[-k_1
t^n\right]$, where $\chi$ is the helix fraction. They obtained the
exponent $n$ close to unity, suggesting one-dimensional growth from
predetermined nuclei and supporting the first order kinetics ideas.
Early optical rotation studies confirmed that simple exponential
kinetic occurs in dilute solutions, but only in the initial stages
of renaturation \cite{djab83,dura85}. Most recently, Guo {\it et
al.~}\cite{guo03} studied initial renaturation rates of gelatin
solutions at semidilute concentrations up to $0.12$\,g/ml and
observed what appeared as a combination of first order and second
order kinetics, i.e. the rate of growth of the normalized helix
fraction had a linear concentration dependence. This seems unusual,
in view of the triple nature of tertiary helix linkages in collagen.
However, the observations of second-order gelation kinetics, in
particular the elastic modulus proportional to $c^2$, have been made
for over a century \cite{veis64}. A two-step mechanism with the rate
limiting step formation of a nucleus of two helices wrapped
together, followed by rapid subsequent wrapping of another coil
segment to form the triple helix has been proposed a long time ago
to account for these findings \cite{coopes70,normand00}. However, we
shall argue that there is a big difference between optical rotation
and rheological studies. The latter return the viscoelastic response
of the gel and thus rely on the triple-helix linkages of the
network. Our opinion \cite{gornall06} is that the properly measured
optical rotation signal is determined by secondary helices and not
the tertiary structure. We attribute the second-order kinetics of
Guo {\it et al.} \cite{guo03} to the range of examined
concentrations - only up to $\sim 0.1$\,g/ml (which is also the
range in the rheological work of Normand {\it et al.}
\cite{normand00}). Overall, examining the literature of the last 30
years, one finds no universality, nor agreement between different
concepts.

\begin{figure} 
\centering \resizebox{0.46\textwidth}{!}{\includegraphics{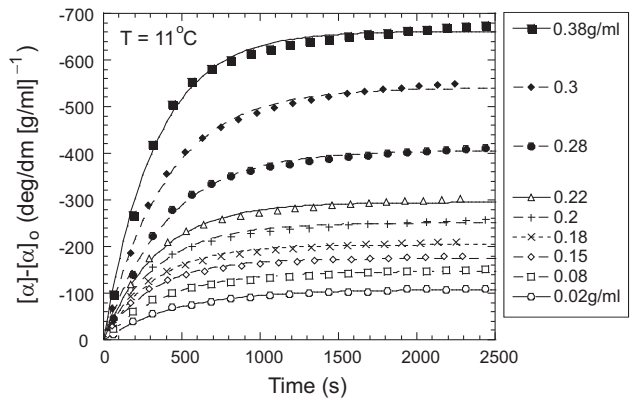}}
\caption{Typical traces of normalized optical rotation, directly
proportional to the secondary helix density, as function of time
after quenching the denatured collagen solution to $T_{\rm
Q}=11^{\rm o}$C. Different data sets correspond to concentrations
labelled on graph. Solid/dashed lines are fits with
$A(1-\exp[-t/\tau])$. } \label{fig:quench}
\end{figure}

In an attempt to link together all of the past findings, we
studied the helix-coil transition in gelatin over a much wider
concentration range, up to $0.4$\,g/ml. The universal master
curves we construct from the new superposition procedure give
explicit predictions for $c \approx 0.65$\,g/ml, and can be
further extended to the full range of temperatures and
concentrations. The high-sensitivity differential optical rotation
detector we used to measure the secondary helix content has been
described in, e.g., \cite{courty06}. The raw measurement of the
total angle of polarization plane rotation, $\Psi$, is divided by
the (constant) sample thickness to produce the rotation rate. It
is then normalized by the solution concentration to obtain a
specific rotation $[\alpha]=(1/c)[d\Psi/dz]$, from which we
subtract a bare value $[\alpha]_0$ corresponding to the average
aminoacid optical activity, separately measured in the coil state
at high temperatures. The resulting difference is proportional to
the concentration of correlated helices in the medium. The typical
reading, for a range of concentrations, is shown in
Fig.~\ref{fig:quench} for a solution quenched from $ 50^{\rm o}$C
to the value $T =11^{\rm o}$C in this case. As a result of
$[\alpha]$ normalization, the $y$-axis in Fig.~\ref{fig:quench} is
directly proportional to the helix fraction in the sample.

\begin{figure} 
\centering \resizebox{0.35\textwidth}{!}{\includegraphics{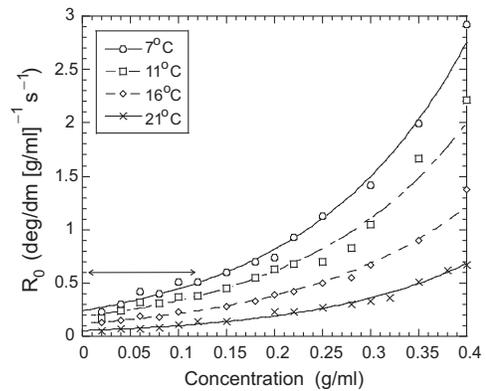}}
\caption{Initial rates of helical nuclei growth, $R_0$ at $t
\rightarrow 0$, as functions of solution concentration, for a
range of quench temperatures labeled on plot. Solid/dashed lines
are fits with exponential $B \,\exp[Y\, c]$. The arrow indicates
the concentration range of earlier studies~\cite{guo03}.}
\label{fig:rates}
\end{figure}

There are important and delicate issues of the slow drift of
collagen towards its natural state~\cite{djab88}, reflected in the
small deviation of the long-time data from the simple exponential
fit of each data set (shown by the lines in Fig.~\ref{fig:quench}).
For instance, the rheological study of Normand {\it et
al.}~\cite{normand00} is heavily based on this regime, where the
tertiary structure (and the gel elasticity) are being consolidated
and the elastic modulus increases dramatically. We, however, are
concerned with the initial rates of secondary helix growth,
$R_0(c,T)$, essentially the slopes $R_0 \equiv d[\alpha ]/dt =
A/\tau $ at $t\rightarrow 0$, from the fitted functions
$A[1-\exp(-t/\tau)]$ in Fig.~\ref{fig:quench}. For each quench
temperature, these slopes depend on the solution concentration. The
summary of this dependence is given in Fig.~\ref{fig:rates}. The
highly non-linear concentration dependence of growth rates is
apparent from the exponential fits of all data sets. We note that
the highest solution concentration quantitatively studied so far was
$c \sim 0.12$~g/ml, by Guo {\it et al.}\cite{guo03}.

\begin{figure} 
\centering \resizebox{0.35\textwidth}{!}{\includegraphics{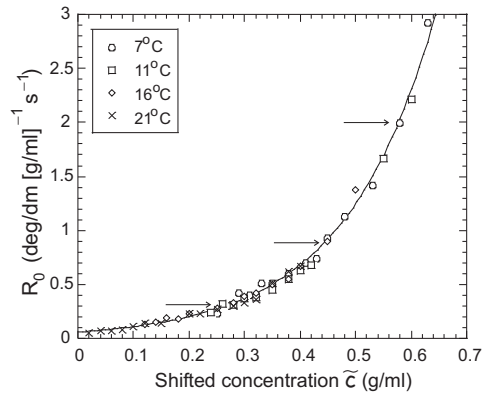}}
\caption{The master curve of growth rate vs. concentration,
$R_0(\widetilde{c})$, superposing the results for different quench
temperatures (labelled on plot). The arrows point at different
data sets; the solid line is a universal fit with $B \,\exp[Y\,
\widetilde{c}])$.} \label{fig:super}
\end{figure}

Guided by the experience in time-temperature and
time-concentration superposition, we notice that the sets of data
for the growth rates  in Fig.~\ref{fig:rates} can be shifted along
the concentration axis by an amount that is a function of quench
temperature. Selecting a reference temperature (at this stage, we
arbitrarily choose $T_{\rm ref}=21^{\rm o}$C) one scales the
concentration $c$ for each data set such that $\widetilde{c}=\beta
\cdot c$, with the coefficient (the shift factor) a function of
the quench temperature for each data set, $\beta = \beta(T)$. The
fact that these sets do superpose means that there is a universal
underlying expression for the growth rate $R_0[\beta(T) c]$. The
resulting master curve, Fig.~\ref{fig:super}, is an indication of
such a universal physical process that controls the helical
nucleation and growth in all regions of the $(T,c)$-phase diagram.

\begin{figure}
\centering \resizebox{0.35\textwidth}{!}{\includegraphics{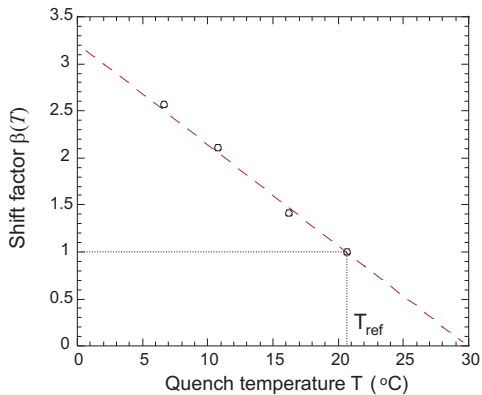}}
\caption{Dependence of the shift factor $\beta$ on quench
temperature $T_{\rm Q}$. The reference temperature chosen at
$21^{\rm o}$C corresponds to the original, non-shifted data -- or
$\beta=1$. The linear fit (dashed line) crosses the $\beta=0$ line
at $\sim 30^{\rm o}$C.} \label{fig:shift}
\end{figure}

It is important to establish a law of the shift factors $\beta$
dependence on temperature. Figure~\ref{fig:shift} gives the plot
of a set $\beta(T)$ required to produce the superposed master
curve for the initial growth rate. It follows quite an obvious
linear function, $\beta = 0.1(T_{\rm c}-T)$, with the ``critical
temperature'' $T_{\rm c}\approx 30^{\rm o}$C (more precisely,
$302.5^{\rm o}$K) defined as the point at which $\beta = 0$. This
temperature is labelled in the gelatin phase diagram and the
sequence of DSC scans, Fig.~\ref{fig:phase}(a). Clearly, its
relation to the transition line is not accidental. In fact we
should conclude that (within errors of our experiment and
analysis) this critical temperature is indeed the line of the
helix-coil transition, $T_{\rm c}=T_{\rm m}(c)$. With this
assumption, the helical growth rate, as obtained from the master
curve in Fig.~\ref{fig:shift}, follows the equation
\begin{equation}
\left. \frac{d[\alpha ]}{dt} \right|_{t\rightarrow 0} = B \, \exp
\left[b(T_{\rm m}-T ) \, c \right] \label{eq:rate}
\end{equation}
 with fixed parameters $B \approx 0.06 \, \hbox{deg/dm
(g/ml)}^{-1}\hbox{s}^{-1}$ and $b \approx 0.6 \,
\hbox{(ml/g)}/^{\rm o}\hbox{K}$. Note that we find no critical or
singular behavior near the transition. The growth rate becomes
exponentially small at $T > T_{\rm m}$. In retrospect, it is
perhaps not surprising that we did not obtain a critical vanishing
of the growth rate, as would be the case with classical phase
transitions, or was suggested by the old Eq.(\ref{eq:flory}). Our
present finding supports the idea of a diffuse nature of
helix-coil transformation in a single chain, with chain
fluctuations capable of creating a non-vanishing helical fraction
above $T_{\rm m}(c)$ and disrupt thermodynamically equilibrium
helices formed below this temperature. In other words, there is no
sharp phase boundary $T_{\rm m}(c)$ for the helix-coil transition.
In equilibrium models of this transition one identifies $T_{\rm
m}$ as a point of steepest gradient in helical fraction $\chi(T)$,
while in our study of kinetics we obtain roughly the same value at
a point when $R_0(c,T)$ changes from the decaying to the growing
exponential function $ \exp[Yc]$.

\begin{figure}[t]
\centering \resizebox{0.35\textwidth}{!}{\includegraphics{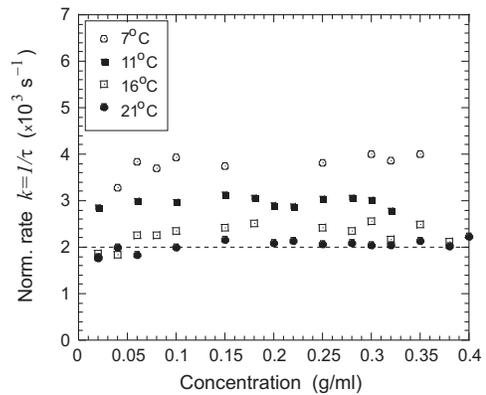}}
\caption{Normalized folding rates $k$ as functions of
concentration at different quench temperatures (labelled on
plot).} \label{fig:k1}
\end{figure}

Having established the universal master curve for the folding rates
$R_0(c,T)$, we must now make contact with the other kinetic studies,
that traditionally focus on the normalized helical fraction, in our
notation defined as
 \begin{equation}
 \chi(t) = \frac{[\alpha ] - [\alpha]_0}{A(c,T)} \
 =1-\exp(-t/\tau) \ ,
 \end{equation}
according to Fig.~\ref{fig:quench} and earlier work \cite{god78}.
The origin of the exponential concentration dependence,
Eq.~(\ref{eq:rate}), is in the constant $A(c)$, while the normalized
rate $k = 1/\tau \equiv R_0/A$. The concentration dependence of this
rate is a signature of process topology, e.g, a constant value means
first-order (single-molecule) folding process. Guo {\it et
al.~}\cite{guo03} claimed a second-order process, i.e.~a linear
dependence $k = k_1 + k_2 c + ...$, but we clearly arrive at
different conclusions. Figure~\ref{fig:k1} shows our results for the
normalized rates $k$ over a concentration range much wider than in
any previous work. There are some delicate issues of fitting the
data that includes the slow long-time tail, which certainly accounts
for the large noise in the $k(c)$ data. However, the qualitative
picture clearly suggests that there is no relevant concentration
dependence and, therefore, the single-chain folding process
(first-order) $k_1$ is in fact dominant at early stages of
renaturation.  This may appear a surprising result as, with
increasing concentration, intramolecular rearrangements of single
chains become limited. Also, the triple-helical nature of collagen
helices is well established indeed \cite{veis64}. However, let us
remember that our experimental method inherently looks at the
secondary helices (the signal from triple helical structures is very
low due to the large wavelength mismatch). The observed first-order
(i.e. single-chain) nature of the principal folding process appears
to be an indication that the transition from coil to secondary helix
is a single chain process, while the association of these helices to
form the triple network linkages occurs subsequently and contributes
less to the helical onset than was previously thought.

In conclusion, by applying a new concentration-temperature
superposition we found a universal master curve describing the
initial folding rates $R_0$ of gelatin solutions over a broad range
of concentrations and temperatures, spanning the nominal helix-coil
transition point $T_m$. The exponential $(c,T)$-dependence of $R_0$
arises from the increasing overall amount of helices in different
systems. In contrast, the normalized helical fraction $\chi(t)$
grows with a $c$-independent rate, suggesting the first-order
kinetics. One needs to study other oligomerizing proteins to test
the further universality of the discovered master curves.

We acknowledge useful discussions with R. Colby, P. Barker, P.
Cicuta and A. Craig. This research was supported by EPSRC.


\begin{thebibliography}{26}
\expandafter\ifx\csname
natexlab\endcsname\relax\def\natexlab#1{#1}\fi
\expandafter\ifx\csname bibnamefont\endcsname\relax
  \def\bibnamefont#1{#1}\fi
\expandafter\ifx\csname bibfnamefont\endcsname\relax
  \def\bibfnamefont#1{#1}\fi
\expandafter\ifx\csname citenamefont\endcsname\relax
  \def\citenamefont#1{#1}\fi
\expandafter\ifx\csname url\endcsname\relax
  \def\url#1{\texttt{#1}}\fi
\expandafter\ifx\csname
urlprefix\endcsname\relax\def\urlprefix{URL }\fi
\providecommand{\bibinfo}[2]{#2}
\providecommand{\eprint}[2][]{\url{#2}}

\bibitem[{\citenamefont{Ferguson and Ferscht}(2003)}]{fersht03}
\bibinfo{author}{\bibfnamefont{N.}~\bibnamefont{Ferguson}} \bibnamefont{and}
  \bibinfo{author}{\bibfnamefont{A.~R.} \bibnamefont{Ferscht}},
  \bibinfo{journal}{Curr. Opin. Struct. Biol.} \textbf{\bibinfo{volume}{13}},
  \bibinfo{pages}{75} (\bibinfo{year}{2003}).

\bibitem[{\citenamefont{Kubelka et~al.}(2004)\citenamefont{Kubelka, Hofrichter,
  and Eaton}}]{kubel04}
\bibinfo{author}{\bibfnamefont{J.}~\bibnamefont{Kubelka}},
  \bibinfo{author}{\bibfnamefont{J.}~\bibnamefont{Hofrichter}},
  \bibnamefont{and} \bibinfo{author}{\bibfnamefont{W.~A.} \bibnamefont{Eaton}},
  \bibinfo{journal}{Curr. Opin. Struct. Biol.} \textbf{\bibinfo{volume}{14}},
  \bibinfo{pages}{76} (\bibinfo{year}{2004}).

\bibitem[{\citenamefont{Elcock and McCammon}(2001)}]{elco01}
\bibinfo{author}{\bibfnamefont{A.~H.} \bibnamefont{Elcock}} \bibnamefont{and}
  \bibinfo{author}{\bibfnamefont{J.~A.} \bibnamefont{McCammon}},
  \bibinfo{journal}{Proc. Nat. Acad. Sci.} \textbf{\bibinfo{volume}{98}},
  \bibinfo{pages}{2990} (\bibinfo{year}{2001}).

\bibitem[{\citenamefont{Zimm and Bragg}(1959)}]{zimm59}
\bibinfo{author}{\bibfnamefont{B.~H.} \bibnamefont{Zimm}} \bibnamefont{and}
  \bibinfo{author}{\bibfnamefont{J.~K.} \bibnamefont{Bragg}},
  \bibinfo{journal}{J. Chem. Phys.} \textbf{\bibinfo{volume}{11}},
  \bibinfo{pages}{526} (\bibinfo{year}{1959}).

\bibitem[{\citenamefont{Ferry}(1970)}]{ferry70}
\bibinfo{author}{\bibfnamefont{J.~D.} \bibnamefont{Ferry}},
  \emph{\bibinfo{title}{Viscoelastic Properties of Polymers (2nd ed.)}}
  (\bibinfo{publisher}{Wiley, NY}, \bibinfo{year}{1970}).

\bibitem[{\citenamefont{Williams et~al.}(1955)\citenamefont{Williams, Landel,
  and Ferry}}]{wlf55}
\bibinfo{author}{\bibfnamefont{M.~L.} \bibnamefont{Williams}},
  \bibinfo{author}{\bibfnamefont{R.~F.} \bibnamefont{Landel}},
  \bibnamefont{and} \bibinfo{author}{\bibfnamefont{J.~D.} \bibnamefont{Ferry}},
  \bibinfo{journal}{J. Amer. Chem. Soc.} \textbf{\bibinfo{volume}{77}},
  \bibinfo{pages}{3701} (\bibinfo{year}{1955}).

\bibitem[{\citenamefont{Trappe and Weitz}(2000)}]{trap00}
\bibinfo{author}{\bibfnamefont{V.}~\bibnamefont{Trappe}} \bibnamefont{and}
  \bibinfo{author}{\bibfnamefont{D.~A.} \bibnamefont{Weitz}},
  \bibinfo{journal}{Phys. Rev. Lett.} \textbf{\bibinfo{volume}{85}},
  \bibinfo{pages}{449} (\bibinfo{year}{2000}).

\bibitem[{\citenamefont{Cicuta et~al.}(2003)\citenamefont{Cicuta, Stancik, and
  Fuller}}]{cicu03}
\bibinfo{author}{\bibfnamefont{P.}~\bibnamefont{Cicuta}},
  \bibinfo{author}{\bibfnamefont{E.~J.} \bibnamefont{Stancik}},
  \bibnamefont{and} \bibinfo{author}{\bibfnamefont{G.~G.}
  \bibnamefont{Fuller}}, \bibinfo{journal}{Phys. Rev. Lett.}
  \textbf{\bibinfo{volume}{90}}, \bibinfo{pages}{236101}
  (\bibinfo{year}{2003}).

\bibitem{gornall06}
\bibinfo{author}{\bibfnamefont{S.}~\bibnamefont{Courty}},
  \bibinfo{author}{\bibfnamefont{J.~L.} \bibnamefont{Gornall}},
  \bibnamefont{and} \bibinfo{author}{\bibfnamefont{E.~M.}
  \bibnamefont{Terentjev}}, \bibinfo{journal}{Biophys. J.}
  \textbf{\bibinfo{volume}{90}}, \bibinfo{pages}{1019}
  (\bibinfo{year}{2006}).

\bibitem{winter86}
  \bibinfo{author}{\bibfnamefont{H.~H.} \bibnamefont{Winter}},
  \bibnamefont{and} \bibinfo{author}{\bibfnamefont{F.}
  \bibnamefont{Chambon}}, \bibinfo{journal}{J. Rheol.}
  \textbf{\bibinfo{volume}{30}}, \bibinfo{pages}{367}
  (\bibinfo{year}{1986}).

\bibitem{goldbart96}
\bibinfo{author}{\bibfnamefont{P.~M.}~\bibnamefont{Goldbart}},
  \bibinfo{author}{\bibfnamefont{H.~E.} \bibnamefont{Castillo}},
  \bibnamefont{and} \bibinfo{author}{\bibfnamefont{A.}
  \bibnamefont{Zippelius}}, \bibinfo{journal}{Adv. Phys.}
  \textbf{\bibinfo{volume}{45}}, \bibinfo{pages}{393}
  (\bibinfo{year}{1996}).

\bibitem{normand00}
\bibinfo{author}{\bibfnamefont{V.}~\bibnamefont{Normand}},
  \bibinfo{author}{\bibfnamefont{S.} \bibnamefont{Muller}},
  \bibinfo{author}{\bibfnamefont{J.-C.} \bibnamefont{Ravey}},
  \bibnamefont{and} \bibinfo{author}{\bibfnamefont{A.}
  \bibnamefont{Parker}}, \bibinfo{journal}{Macromolecules}
  \textbf{\bibinfo{volume}{33}}, \bibinfo{pages}{1063}
  (\bibinfo{year}{2000}).

\bibitem[{\citenamefont{Ramachandran}(1967)}]{rama67}
\bibinfo{author}{\bibfnamefont{G.~N.} \bibnamefont{Ramachandran}},
  \emph{\bibinfo{title}{Treatise on Collagen}} (\bibinfo{publisher}{N. Y. Acad.
  Press}, \bibinfo{year}{1967}).

\bibitem[{\citenamefont{Pezron et~al.}(1990)\citenamefont{Pezron, Djabourov,
  Bosio, and Leblond}}]{pezr90}
\bibinfo{author}{\bibfnamefont{I.}~\bibnamefont{Pezron}},
  \bibinfo{author}{\bibfnamefont{M.}~\bibnamefont{Djabourov}},
  \bibinfo{author}{\bibfnamefont{L.}~\bibnamefont{Bosio}}, \bibnamefont{and}
  \bibinfo{author}{\bibfnamefont{J.}~\bibnamefont{Leblond}},
  \bibinfo{journal}{J. Polym. Sci., Part B: Polym. Phys.}
  \textbf{\bibinfo{volume}{28}}, \bibinfo{pages}{1823} (\bibinfo{year}{1990}).

\bibitem[{\citenamefont{Djabourov et~al.}(1993)\citenamefont{Djabourov, Bonnet,
  Kaplan, Favard, Favard, Lechaire, and Maillard}}]{djab93}
\bibinfo{author}{\bibfnamefont{M.}~\bibnamefont{Djabourov}},
  \bibinfo{author}{\bibfnamefont{N.}~\bibnamefont{Bonnet}},
  \bibinfo{author}{\bibfnamefont{H.}~\bibnamefont{Kaplan}},
  \bibinfo{author}{\bibfnamefont{N.}~\bibnamefont{Favard}},
  \bibinfo{author}{\bibfnamefont{P.}~\bibnamefont{Favard}},
  \bibinfo{author}{\bibfnamefont{J.}~\bibnamefont{Lechaire}}, \bibnamefont{and}
  \bibinfo{author}{\bibfnamefont{M.}~\bibnamefont{Maillard}},
  \bibinfo{journal}{J. Phys. II (Paris)} \textbf{\bibinfo{volume}{3}},
  \bibinfo{pages}{611} (\bibinfo{year}{1993}).

\bibitem[{\citenamefont{Harrington and Rao}(1970)}]{harr70}
\bibinfo{author}{\bibfnamefont{W.~F.} \bibnamefont{Harrington}}
  \bibnamefont{and} \bibinfo{author}{\bibfnamefont{N.~V.} \bibnamefont{Rao}},
  \bibinfo{journal}{Biochemistry} \textbf{\bibinfo{volume}{9}},
  \bibinfo{pages}{3714} (\bibinfo{year}{1970}).

\bibitem[{\citenamefont{Guo et~al.}(2003)\citenamefont{Guo, Colby, Lusignan,
  and Howe}}]{guo03}
\bibinfo{author}{\bibfnamefont{L.}~\bibnamefont{Guo}},
  \bibinfo{author}{\bibfnamefont{R.~H.} \bibnamefont{Colby}},
  \bibinfo{author}{\bibfnamefont{C.~P.} \bibnamefont{Lusignan}},
  \bibnamefont{and} \bibinfo{author}{\bibfnamefont{A.~M.} \bibnamefont{Howe}},
  \bibinfo{journal}{Macromolecules} \textbf{\bibinfo{volume}{36}},
  \bibinfo{pages}{10009} (\bibinfo{year}{2003}).

\bibitem[{\citenamefont{Goddard et~al.}(1978)\citenamefont{Goddard, Biebuyck,
  Daumerie, Naveau, and Mercier}}]{god78}
\bibinfo{author}{\bibfnamefont{P.}~\bibnamefont{Goddard}},
  \bibinfo{author}{\bibfnamefont{J.~J.} \bibnamefont{Biebuyck}},
  \bibinfo{author}{\bibfnamefont{M.}~\bibnamefont{Daumerie}},
  \bibinfo{author}{\bibfnamefont{H.}~\bibnamefont{Naveau}}, \bibnamefont{and}
  \bibinfo{author}{\bibfnamefont{J.~P.} \bibnamefont{Mercier}},
  \bibinfo{journal}{J. Polym. Sci., Polym. Phys. Edn.}
  \textbf{\bibinfo{volume}{16}}, \bibinfo{pages}{1817} (\bibinfo{year}{1978}).

\bibitem[{\citenamefont{Bohidar et~al.}(1993)\citenamefont{Bohidar and Jena}}]{bohi93}
\bibinfo{author}{\bibfnamefont{H.~B.}~\bibnamefont{Bohidar}},
  \bibinfo{author}{\bibfnamefont{S.~S.} \bibnamefont{Jena}},
  \bibinfo{journal}{J. Chem. Phys.}
  \textbf{\bibinfo{volume}{98}}, \bibinfo{pages}{8970} (\bibinfo{year}{1993}).

\bibitem[{\citenamefont{Norton et~al.}(1983)\citenamefont{Norton, Goodall,
  Morris, and Rees}}]{nort83}
\bibinfo{author}{\bibfnamefont{I.~T.} \bibnamefont{Norton}},
  \bibinfo{author}{\bibfnamefont{D.~M.} \bibnamefont{Goodall}},
  \bibinfo{author}{\bibfnamefont{E.~R.} \bibnamefont{Morris}},
  \bibnamefont{and} \bibinfo{author}{\bibfnamefont{D.~A.} \bibnamefont{Rees}},
  \bibinfo{journal}{J. Chem. Soc. Fraraday Trans. (I)}
  \textbf{\bibinfo{volume}{79}}, \bibinfo{pages}{2489} (\bibinfo{year}{1983}).

\bibitem[{\citenamefont{Djabourov et~al.}(1988)\citenamefont{Djabourov,
  Leblond, and Papon}}]{djab88}
\bibinfo{author}{\bibfnamefont{M.}~\bibnamefont{Djabourov}},
  \bibinfo{author}{\bibfnamefont{J.}~\bibnamefont{Leblond}}, \bibnamefont{and}
  \bibinfo{author}{\bibfnamefont{P.}~\bibnamefont{Papon}}, \bibinfo{journal}{J.
  Phys. France} \textbf{\bibinfo{volume}{49}}, \bibinfo{pages}{319}
  (\bibinfo{year}{1988}).

\bibitem[{\citenamefont{Flory and Weaver}(1960)}]{flor60}
\bibinfo{author}{\bibfnamefont{P.~J.} \bibnamefont{Flory}} \bibnamefont{and}
  \bibinfo{author}{\bibfnamefont{E.~S.} \bibnamefont{Weaver}},
  \bibinfo{journal}{J. Am. Chem. Soc.} \textbf{\bibinfo{volume}{82}},
  \bibinfo{pages}{4518} (\bibinfo{year}{1960}).

\bibitem[{\citenamefont{TerMeer et~al.}()\citenamefont{TerMeer, Lips, and
  Busnel}}]{term90}
\bibinfo{author}{\bibfnamefont{H.~U.} \bibnamefont{TerMeer}},
  \bibinfo{author}{\bibfnamefont{A.}~\bibnamefont{Lips}}, \bibnamefont{and}
  \bibinfo{author}{\bibfnamefont{J.-P.} \bibnamefont{Busnel}}, in
  \emph{\bibinfo{booktitle}{Physical networks: polymers and gels}}, edited by
  \bibinfo{editor}{\bibfnamefont{W.}~\bibnamefont{Burchard}} \bibnamefont{and}
  \bibinfo{editor}{\bibfnamefont{S.}~\bibnamefont{Ross-Murphy}} (Elselvier, 1990).

\bibitem[{\citenamefont{Avrami}(1941)}]{avr41}
\bibinfo{author}{\bibfnamefont{M.}~\bibnamefont{Avrami}}, \bibinfo{journal}{J.
  Chem. Phys.} \textbf{\bibinfo{volume}{9}}, \bibinfo{pages}{177}
  (\bibinfo{year}{1941}).

\bibitem[{\citenamefont{Djabourov and Papon}(1983)}]{djab83}
\bibinfo{author}{\bibfnamefont{M.}~\bibnamefont{Djabourov}} \bibnamefont{and}
  \bibinfo{author}{\bibfnamefont{P.}~\bibnamefont{Papon}},
  \bibinfo{journal}{Polymer} \textbf{\bibinfo{volume}{24}},
  \bibinfo{pages}{537} (\bibinfo{year}{1983}).

\bibitem[{\citenamefont{Durand et~al.}(1985)\citenamefont{Durand, Emry, and
  Chatelier}}]{dura85}
\bibinfo{author}{\bibfnamefont{D.}~\bibnamefont{Durand}},
  \bibinfo{author}{\bibfnamefont{J.~R.} \bibnamefont{Emry}}, \bibnamefont{and}
  \bibinfo{author}{\bibfnamefont{J.~Y.} \bibnamefont{Chatelier}},
  \bibinfo{journal}{Int. J. Biol. Macromol.} \textbf{\bibinfo{volume}{7}},
  \bibinfo{pages}{315} (\bibinfo{year}{1985}).

\bibitem{veis64}
\bibinfo{author}{\bibfnamefont{A.} \bibnamefont{Veis}},
  \emph{\bibinfo{title}{Macromolecular Chemistry of Gelatin}}
  (\bibinfo{publisher}{Academic Press, NY}, \bibinfo{year}{1964}).

\bibitem{coopes70}
  \bibinfo{author}{\bibfnamefont{I.~H.} \bibnamefont{Coopes}},
  \bibinfo{journal}{J. Polym. Sci. A-1} \textbf{\bibinfo{volume}{8}},
  \bibinfo{pages}{1793} (\bibinfo{year}{1970}).

\bibitem[{\citenamefont{Courty et~al.}(2006)\citenamefont{Courty, Tajbakhsh,
  and Terentjev}}]{courty06}
\bibinfo{author}{\bibfnamefont{S.}~\bibnamefont{Courty}},
  \bibinfo{author}{\bibfnamefont{A.~R.} \bibnamefont{Tajbakhsh}},
  \bibnamefont{and} \bibinfo{author}{\bibfnamefont{E.~M.}
  \bibnamefont{Terentjev}}, \bibinfo{journal}{Phys. Rev. E}
  \textbf{\bibinfo{volume}{73}},  \bibinfo{pages}{011803} (\bibinfo{year}{2006}).

\end{thebibliography}

\end{document}